\documentclass[prl,floats,twocolumn,showpacs,floatfix]{revtex4}
\usepackage[dvips]{epsfig}

\begin{document}
\title{Roughness induced boundary slip in microchannel flows}
\author{Christian Kunert}
\author{Jens Harting}

\pacs{83.50.Rp,68.08.-p}

\date{\today}
\affiliation{Institute for Computational Physics, University of Stuttgart, Pfaffenwaldring 27, D-70569 Stuttgart, Germany}

\date{\today}

\begin{abstract}
Surface roughness becomes relevant if typical length scales
of the system are comparable to the scale of the variations as it is the
case in microfluidic setups. Here, an apparent boundary slip is often
detected which can have its origin in the assumption of perfectly smooth
boundaries. We investigate the problem by means of lattice Boltzmann (LB)
simulations and introduce an ``effective no-slip plane'' at an
intermediate position between peaks and valleys of the surface.
Our simulations show good agreement with
analytical results for sinusoidal boundaries, but can be extended to
arbitrary geometries and experimentally obtained surface data. We find
that the detected apparent slip is independent of the detailed boundary 
shape, but only given by the distribution of surface heights. Further, we
show that the slip diverges as the amplitude of the roughness increases.
\end{abstract}
\maketitle

In microfluidic systems the surface to volume ratio is large causing
boundary effects to be significantly more important than in macroscopic
devices. Since even on atomic or
molecular scales a perfectly smooth surface is an idealized model, the
shape of the boundary is an important property.
Additionally, it is of technological interest to design surfaces with well
defined structures and properties~\cite{stroock-etal-02,joseph-etal-06}.
A commonly investigated surface property is the apparent slip originating
for example from the surface wettability, electrostatic interactions,
impurities, or surface structuring~\cite{neto-etal-05}.
Navier characterized hydrodynamic slip by postulating that the fluid
velocity $v(x)$ at the boundary $(x=0)$ is proportional to the shear rate
$\frac{\partial v}{\partial x}$ and the slip length $\beta$~\cite{navier-23}.
For macroscopic systems the simple no-slip boundary condition ($\beta=0$)
is a valid assumption. However, if the height of surface variations is not
small compared to typical length scales of the system, the position of
the boundary is not clearly defined and experiments might detect slip
due to not accurately determined wall positions.
%
%
The influence of roughness on the slip length $\beta$ has
been investigated by numerous authors. Roughness leads to higher drag
forces and thus to no slip on macroscopic scales, as shown by
Richardson~\cite{richardson-73} and  Jansons~\cite{jansons-87}.
This was  experimentally demonstrated by McHale
and Newton~\cite{mchale-newton-04}. 
Jabbarzadeh et al.~performed molecular dynamics
(MD) simulations of Couette flow between sinusoidal walls and found that
slip appears for roughness amplitudes smaller than the molecular length
scale~\cite{jabbarzadeh-etal-00}. Also, roughness can cause
pockets to be filled with vapor or gas nano bubbles leading to apparent
slip~\cite{du-goubaidoulline-johansmann-04,joseph-etal-06}.  
Recently, Sbragaglia et al. applied the lattice Boltzmann (LB) method to
simulate fluids in the vicinity of microstructured hydrophobic
surfaces~\cite{sbragaglia-06} and Varnik et
al.~\cite{varnik-dorner-raabe-06} have shown 
that even in small geometries rough channel
surfaces can cause flow to become turbulent. A common setup to measure
slip is to utilize a modified atomic force microscope (AFM) to
oscillate a colloidal sphere in the vicinity of a
boundary~\cite{bib:vinogradova-95,bonaccurso-03,vinogradova-yakubov-06}.
Vinogradova and Yakubov demonstrated that assuming a wrong position of
the surface during measurements can lead to substantial errors
in the determined slip lengths~\cite{vinogradova-yakubov-06}. They
showed that measurements can be interpreted by assuming a modified sphere
radius instead of Navier's slip condition, so that
the position of a no slip wall would be between peaks and valleys
of the rough surface. In this paper we follow this idea. We
answer the question at which distinct position the ``effective boundary''
has to be placed and study the influence of a wrongly determined wall
position numerically.

Panzer et al.~gave an analytical equation for $\beta$ for small cosine-shaped
surface variations~\cite{panzer-liu-einzel-92}. It is applicable to two
infinite planes separated by a distance $2d$ being much larger than the highest
peaks $h_{\rm max}$. Surface variations are determined by peaks of height
$h_{\rm max}$, valleys at $h_{\rm min}$ and given by $h(z)=h_{\rm max}/2+h_{\rm
max}/2\cos(qz)$. Here, $q$ is the wave number and the corresponding slip length
is found to be 
\begin{equation}
\label{eq:panzer}
\beta=\frac{-h_{\rm max}}{2}\left(1+ k\frac{1-\frac{1}{4}k^2+\frac{19}{64}k^4+\mathcal{O}(k^6) }{1+k^2(1-\frac{1}{2}k^2)+\mathcal{O}(k^6)}\right).
\end{equation}
Higher order terms cannot easily be calculated analytically and are neglected.
Thus, Eq.~\ref{eq:panzer} is valid only for $k=qh_{\rm max}/2\ll 1$. However,
for realistic surfaces, $k$ can become substantially larger
than $1$ causing the theoretical approach to fail. Here, only numerical
simulations can be applied to describe arbitrary boundaries.

\begin{figure}[h]
\centerline{\hspace{1.0cm}\epsfig{file=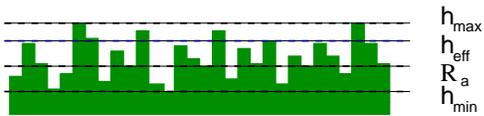,width=0.85\linewidth}}
\caption{ \label{fig:hydrodynamicwall} (Color online) The effective 
boundary height $h_{\rm eff}$ is found between the deepest valley at $h_{\rm
min}$ and the highest peak at $h_{\rm max}$.
}
\end{figure}

\begin{figure}[h]
\centerline{\epsfig{file=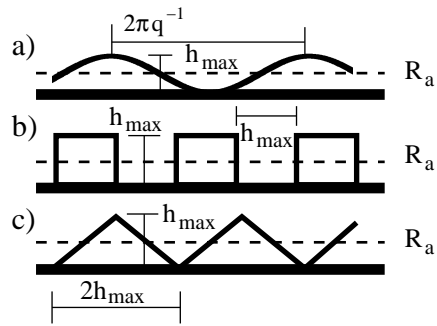,angle=0,width=0.65 \linewidth}}
\caption{\label{fig:profile}
Periodic surfaces: a)
cosines given by $h(x)=h_{\rm max}/2+h_{\rm max}/2\cos(qx)$. b) squares with
 height and separation given by $h_{\rm max}$. c) triangles, 
$h_{\rm max}$ high and $2h_{\rm max}$ wide.}
\end{figure}

We use a 3D LB model as presented
in~\cite{succi-01,bib:jens-harvey-chin-venturoli-coveney:2005,harting-kunert-herrmann-06}
to simulate pressure driven flow between two infinite rough walls. Previously,
we applied the method to study flows of simple fluids and complex mixtures containing
surfactant in hydrophobic
microchannels~\cite{harting-kunert-herrmann-06,bib:jens-kunert:2007a}. Here, we
only shortly describe our method and refer to the literature
for details. The lattice Boltzmann equation, $\eta_i({\bf x}+{\bf c}_i,
t+1) - \eta_i({\bf x},t) = \Omega_i$, with $i= 0,1,\dots,b$, describes the time
evolution of the single-particle distribution $\eta_i({\bf x},t)$,
indicating the amount of quasi particles with velocity ${\bf c}_i$, at site
${\bf x}$ on a 3D lattice of coordination number $b=19$, at time-step $t$.
We choose the Bhatnagar-Gross-Krook collision operator
 $\Omega_i =
 -\tau^{-1}(\eta_i({\bf x},t) - \eta_i^{
\, {\rm eq}}({\bf u}({\bf x},t),\eta({\bf x},t)))$,
with mean collision time $\tau$ and equilibrium
distribution $\eta_i^{\rm eq}$~\cite{harting-kunert-herrmann-06,succi-01}.
Simulation lattices are 256 lattice units long in flow direction and the planes
are separated by 62 sites. Periodic boundary conditions are imposed in the
remaining direction allowing us to keep the resolution as low as 16 lattice
units. A pressure gradient is obtained as described
in~\cite{harting-kunert-herrmann-06}.
An effective boundary position can be found by fitting the parabolic
flow profile
\begin{equation}
\label{eq:profil02}
v_z(x)=\frac{1}{2 \mu}\frac{\partial P}{\partial z}
\left[ d^2-x^2-2d\beta \right]
\end{equation}
via the distance $2d=2d_{\rm eff}$.
$\beta$ is set to 0 here and viscosity $\mu$ as well as pressure
gradient $\frac{\partial P}{\partial z}$ are given by the simulation. 
To obtain an average value for $d_{\rm eff}$, a sufficient number of
individual profiles at different positions $z$ are taken into account.
Alternatively, the mass flow $\int v(x) \rho\, {\rm d}x$ can be computed to obtain $2d_{\rm eff}$.
Both methods are equivalent and produce identical results. The so found
$d_{\rm eff}$ gives the position of the effective
boundary and the effective height $h_{\rm eff}$ of the rough surface is then
defined by $d-d_{\rm eff}$ (see Fig. \ref{fig:hydrodynamicwall}).
As rough model surfaces we choose a randomly generated roughness and three
periodic ones for which the average height or average roughness $R_a$ is given
by $h_{\rm max}/2$. Cosine-shaped boundaries are given by $h(x)=h_{\rm
max}/2+h_{\rm max}/2\cos(qx)$, squares have a height of $h_{\rm max}$ and are
separated by $h_{\rm max}$ lattice sites. Triangular structures are
$2h_{\rm max}$ wide and have a height of $h_{\rm max}$ (see
Fig.~\ref{fig:profile}). 
Randomly generated surface structures are created by choosing for every
lattice position of the boundary the height $h(x)$ as a random
integer number between $0$ and $h_{\rm max}$. For
determining $h_{\rm eff}$ we average over five surfaces generated with
different sequences of uniformly distributed random numbers.
All wall types are geometrically similar, i.e., the effective height $h_{\rm eff}$
scales linearly with $h_{\rm max}$. 
%

\begin{figure}[h]
\centerline{
\epsfig{file=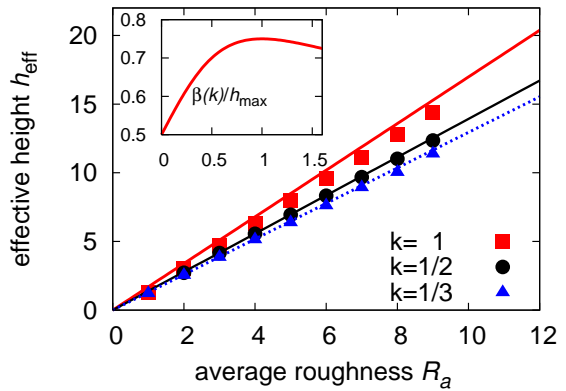,angle=270,width=0.90 \linewidth} }
\caption{ \label{fig:sinus01}
(Color online) Effective height $h_{\rm eff}$ over average roughness $R_a$ for a cosine
geometry and different variations $k$. Symbols denote numerical data and
lines are given by Eq.~\ref{eq:panzer}. The
inset shows $\beta(k)/h_{\rm max}$ according to equation (\ref{eq:panzer}). For
$k>1$ the slope becomes negative, demonstrating that the theory fails for more
complex surface structures.
}
\end{figure}

In Fig.~\ref{fig:sinus01} the effective height $h_{\rm eff}$ obtained
from our simulations is plotted versus $R_a$ for cosine shaped surfaces
with $qh_{\rm max}/2=k=1,\frac{1}{2},\frac{1}{3}$ (symbols). Lines are given
by the analytical solution of Eq.~\ref{eq:panzer}. For $k<1$ the
simulated data agrees within 2.5\% with Panzer's prediction. However,
for $k=1$ a substantial deviation between numerical and analytical
solutions can be observed because Eq.~\ref{eq:panzer} is valid for small
$k$ only. The inset of Fig.~\ref{fig:sinus01} depicts the ratio of
$\beta/h_{\rm max}$ according to the theory of Panzer. 
In the case of large $k>1$, the theory is not able to correctly reproduce
the increase of $\beta$ with increasing $h_{\rm max}$ anymore. Instead,
$\beta/h_{\rm max}$ becomes smaller again due to missing higher order
contributions in Eq.~\ref{eq:panzer}. Our simulations do not suffer from
such limitations allowing us to study arbitrarily complex surface
geometries.

\begin{figure}[h]
\centerline{
\epsfig{file=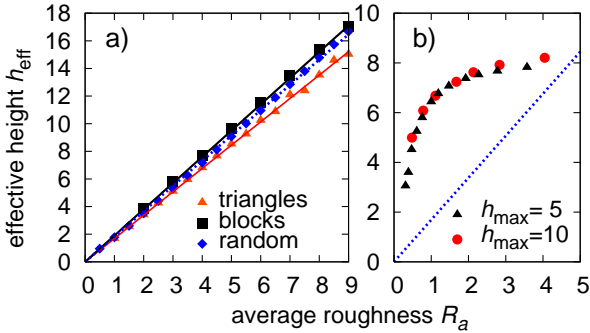,angle=270,width=0.95 \linewidth}
}
\caption{\label{fig:random01}(Color online) a) Effective height $h_{\rm eff}$ versus
$R_a$ for triangles, blocks (see Fig.~\ref{fig:profile}), and an equally
distributed random roughness.
b) $h_{\rm eff}$ versus $R_a$ for
triangles with $h_{\rm max}=5$ and $10$. The distance between triangles
$a$ is varied to obtain the given $R_a$. Values of $h_{\rm max}=5$ are
scaled by a factor of $2$.}
\end{figure}

In Fig. \ref{fig:random01}a $h_{\rm eff}$ is plotted versus $R_a$ for
different types of roughness. By performing a linear fit to the data as
given by the lines we find for the uniformly distributed roughness that
the position of the effective wall is at $c=1.84$ times the average height
of the roughness $R_a=h_{\rm max}/2$ or at 92\% of the maximum height
$h_{\rm max}$. For squares and triangular structures we find constants of
proportionality of $c=1.90$ and $c=1.69$ indicating that the shape of the
surface variations indeed affects the position of the effective boundary.
However, the effect of the shape is small compared the effect of the
height of the variations. All surface structures are geometrically similar causing
the linear dependence between $h_{\rm eff}$ and $R_a=h_{\rm max}/2$.
When converting our 3D random roughness into a purely 2D structure, the
difference in the measured constant of proportionality $c$ is in the range
of the error of the fit algorithm. This is a surprising result since in
three dimensions the flow can pass sidewise a roughness element. The
measured $h_{\rm eff}$ is found to be independent of the flow velocity
over more than three decades and does not depend on the pressure either,
i.e., $h_{\rm eff}$ is independent of the Reynolds number.
 
\begin{figure}[h]
\centerline{
\epsfig{file=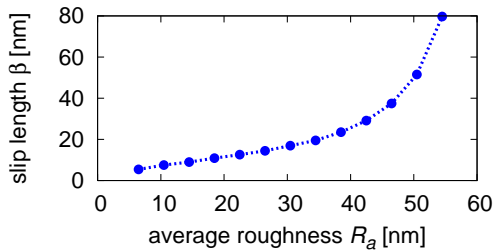,angle=270,width=0.80\linewidth}}
\caption{\label{fig:random02}(Color online)
Slip length $\beta$ versus $R_a$ for water and the randomly distributed roughness. The line is a guide to the eye. By assuming $h_{\rm
eff}=h_{\rm max}$ as it is common in experiments, $\beta$ is in the range of
$h_{\rm max}-h_{\rm eff}$ for small $R_a$, but for larger $R_a$ the apparent slip diverges. 
}
\end{figure}

In reality high pikes on a smooth surface may occur, so that the average
roughness $R_a$ is much smaller than the maximum height $h_{\rm max}$. To
observe such cases we simulate a triangle geometry with additional void
space $a$ between the roughness elements. 
As maximum height $h_{\rm max}$ we
choose 5 and 10 lattice sites. In similarity to Fig.~\ref{fig:random01} we
plot the effective surface height $h_{\rm eff}$ over the average roughness
$R_a$ in Fig.~\ref{fig:random02}. In this case the average roughness is
smaller than the half of $h_{\rm max}$, i.e., $R_a=\frac{ h_{\rm
max}^2}{2h_{\rm max}+a} \le h_{\rm max}/2$.  The values of $h_{\rm max}=5$
are scaled by a factor of 2 to fit them with the values of $h_{\rm
max}=10$. Due to the geometrical similarity of the surface structure this scaling
is possible. For comparison with Fig.~\ref{fig:random01}a the linear fit
with slope $c=1.69$ is plotted. In Fig.~\ref{fig:random01}b we see that
the maximum height has the strongest influence on the effective height
$h_{\rm eff}$ and not the distance $a$. 
For small $R_a$ created by a large
additional distance $a$, $h_{\rm eff}$ converges to zero corresponding to
a flat surface. For small $a$ the data converges to the triangle geometry
as given in Fig.~\ref{fig:random01}a. For a medium $a\approx 2h_{\rm max}$
the effective wall is still in the range of $75\%$ of the maximum height
$h_{\rm max}$. This is an important result, since it demonstrates that the
distance between $h_{\rm eff}$ and $h_{\rm min}$ can be much larger than
$R_a$ and that in such cases $h_{\rm eff}>6\cdot R_a$ can be obtained. On
the other hand, for large $a$ this results in $h_{\rm eff}<0.7 \cdot
h_{\rm max}$. Therefore, in the case of large $a \approx 2 h_{\rm max}$,
the effective wall $h_{\rm eff}$ cannot be approximated by the maximum
height $h_{\rm max}$ nor by the average roughness $R_a$. 

\begin{figure}[h]
\centerline{
\epsfig{file=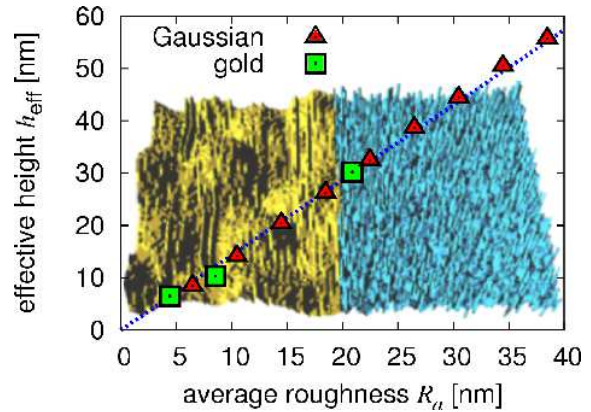,width=0.87 \linewidth}}
\caption{\label{fig:zweifl}(Color online) Simulated $h_{\rm eff}$ versus $R_a$ for a gold
coated glass surface and a randomly generated surface with Gaussian
distributed heights. The background image visualizes the gold surface
(left) and the artifically generated structure (right).
}
\end{figure}

We obtained AFM data of a gold coated glass surface with a maximum peak to
valley distance of 64nm. The sample size is 1${\rm\mu m^2}$ represented by
$512\times 512$ data points. A lattice constant of the LB simulation can
be scaled to 1.9nm by setting the relaxation time $\tau$ to 1.15 and by
mapping the speed of sound and the viscosity to the values for water
($c_{s}=1.5\cdot10^{3}{\rm m/s}$, $\mu=1.02\cdot10^{-6}{\rm m^2/s}$).
$h_{\rm eff}$ can then be measured as in previous paragraphs of this paper
by loading the AFM data onto our simulation lattice. For the simulations
presented in this paragraph, the channel width is set to 128 lattice
units. The simulated effective height of the gold surface is depicted by
the square at $R_a$=21nm in Fig.~\ref{fig:zweifl}. Data points at
$R_a=4$ and $8$ are obtained by downscaling the original data set.
We find that the distribution of surface
heights follows a Gaussian distribution and use this distribution to
generate an artificial random surface with identical height distribution.
In contrast to the AFM data, our data points are fully uncorrelated, while
the gold surface shows distinct structural properties as can be observed
in the background images of Fig.~\ref{fig:zweifl}. For artificial
surfaces, the average roughness $R_a$ can be scaled by scaling the width
of the distribution of random numbers allowing us to determine $h_{\rm
eff}$ for $R_a$ up to 40nm. As shown by the dotted line, the measured
$h_{\rm eff}$ linearly depends on $R_a$ with a constant of proportionality
of $c=1.43$. The data obtained from the gold coated surface follows the same
linear dependence demonstrating that the actual shape of a surface does
not influence the effective surface position, but only the distribution of
heights needs to be known.

The most important question to be answered by our simulations is the
effect of a wrongly assumed position of a surface on experimental
measurements. As mentioned in the introduction many groups use an
approaching method to measure the slip length $\beta$. Here, a colloidal
sphere at the tip of a cantilever immersed in a fluid is oscillated in the
vicinity of a surface, or the two cylinders of a surface force apparatus
(SFA) are brought close to each other. The distance between the surfaces
can become very small -- even down to contact. To study the influence of
the roughness on an apparent slip effect, we assume the surface to be
placed at $h_{\rm max}$ as it is commonly done in
experiments~\cite{bonaccurso-03}. Then, we measure the slip length $\beta$
by fitting Eq.~\ref{eq:profil02}.
The wrong position of the surface causes a substantial error in the detected slip as can
be inferred from Fig.~\ref{fig:random02}. Here, $\beta$ is given versus
$R_a$ for randomly generated boundaries with the heights of the
surface obstacles following the Gaussian distribution given by the AFM
data of the gold surface. For small $R_a$ (and thus
large separation of the plates) $\beta$ is in the range of $h_{\rm max}-h_{\rm
eff}$ and can be neglected in most practical cases. However, the detected slip
diverges if $R_a$ becomes large and grows to 80nm for $R_a=55$nm. Here, a large
$R_a$ is equivalent to the channel width becoming very small -- an effect also
common in typical surface approaching experiments or microchannel flows. For
curved surfaces, as they are utilized in surface force apparatuses or AFM based
slip measurements, the detected $\beta$ can be even larger due to higher order
components of the flow field. This might explain experiments reporting large
slip lengths of $\beta \approx 100 {\rm
nm}$~\cite{lauga-brenner-stone-05,neto-etal-05}.

In conclusion we performed LB simulations of pressure
driven flow between two rough plates. By varying the roughness we found
that there exists an imaginary effective plane where the no slip boundary
condition is valid. We compared our results to analytic calculations
of Panzer et al. and found good agreement in the case of small variations
($k<1$). Large and more realistic perturbations
($k>1$) can only be covered by simulations as presented in this paper. 
By simulating flow of water along a gold coated surface and a randomly
generated one with identical height distribution, we demonstrated that
the position of the effective plane is independent of the actual boundary
structure and that only the distribution of heights is relevant. We
showed that apparent slip due to errornous assumptions of the surface
structure can become very large if the
distance between the boundaries is small -- as it is typical in
dynamic microfluidic experiments. 
Our simulations can be of practical importance for experimental
measurements of boundary slip induced for example by
electrostatic interactions, surface wettability or impurities. 
Due to the precise measurements needed, ignoring the influence of surface
roughness leads to substantial errors in the determined slip. A
simulation of the flow along a surface generated from AFM data allows to
determine how an experimentally detected slip might have to be corrected
in order to take the surface structure into account. 

We thank H.~Gong for the AFM data and O.I.~Vinogradova, M.~Rauscher, and
M.~Hecht for fruitful discussions. This work was financed
within the DFG priority program ``nano- and microfluidics'' and by the
``Landesstiftung Baden-W\"urttemberg''. Computations were performed at
the Neumann Institute for Computing, J\"{u}lich.
 

\end{document}